\documentstyle[preprint,epsf]{jpsj}
\def\runtitle{ 
Electronic States in Half-Filled Correlated System 
}
\def\runauthor{Masahisa {\sc Tsuchiizu} and Yoshikazu {\sc Suzumura}}

\def\td{t_{\rm d}}

\def\d{{\rm d}}
\def\lan{\left\langle}
\def\ran{\right\rangle}
\def\dl{{{\rm d} \over{{\rm d}l}} \,}
\def\dx{{\rm d}x\,}
\def\e{{\rm e}}
\def\i{{\rm i}}
\def\virg{\;\;,}
\def\point{\;\,.}
\def\vf{v_{\rm F}}
\def\kf{k_{\rm F}}

\def\ek{\varepsilon_{k}}

\def\ggs{\buildrel\textstyle > \over {\hbox{\raise0.2ex\hbox{$\sim$}}}}
\def\lls{\buildrel\textstyle < \over {\hbox{\raise0.2ex\hbox{$\sim$}}}}
\def\gsim{\,\lower0.75ex\hbox{$\ggs$}\,}
\def\lsim{\,\lower0.75ex\hbox{$\lls$}\,}
\def\N{\hat{N}}

\def\O{{\cal O}}
\def\too{\mathop{\longrightarrow}}
\newcommand{\br}{r}
\newcommand{\bR}{R}

\def\jo #1#2#3#4{#1 {\bf #2} (#3) #4} 

\def\PRB{Phys.\ Rev.\ B}
\def\PRL{Phys.\ Rev.\ Lett.}

\def\JPF{J.\ Phys.\ France}

\def\RMP{Rev.\ Mod.\ Phys.}
\def\PTP{Prog.\ Theor.\ Phys.}

\def\ADV{Adv.\ Phys.}

\title
{
Electronic States in Half-Filled Correlated System \\
  with Alternating Potential
 }

\author{
Masahisa {\sc Tsuchiizu}$^a$,
   and Yoshikazu  {\sc Suzumura}$^{a,b}$
}

\inst{
$^{a}$Department of Physics, Nagoya University, Nagoya 464-8602 \\ 
$^{b}$CREST, Japan Science and Technology Corporation (JST) \\
}

\recdate{August 16, 1999}

\abst{
    The effect of an alternating potential on a 
     one-dimensional half-filled Hubbard model with  
      repulsive interaction  has been examined 
        by applying the renormalization group method 
         to the bosonized Hamiltonian.  
  The electronic state, which is determined  by the competition between 
      alternating potential and umklapp scattering, is calculated 
  where the relevance and the irrelevance of the alternating potential 
    leads to the band insulator and the Mott insulator respectively. 
 The  excitation gaps for charge and spin fluctuations  are  calculated 
  for both states.  
 }

\kword{charge gap, half-filling,   alternating  potential, 
 umklapp scattering, Mott insulator}

\begin{document}
\sloppy
\maketitle

\section{Introduction} 
  An insulator with mixed ionic-covalent character 
   in the presence  of electron correlation is of interest 
     since  the large  repulsive interaction leads to a Mott insulator 
        due to  umklapp scattering     at half-filling.

  The effect of electron correlation on such an  insulator 
    has been studied  in some materials 
      which consist of two-kinds of atoms per cell.  
  In the  perovskite oxide compounds,  BaTiO$_3$, 
   where the  ferroelectricity   is usually discussed   
     based on  the lattice distortion,   
      the covalency  plays an important role  due to 
    the following fact. 
 The first principle calculation shows that 
    Ba$^{2+}$Ti$^{2.89+}$O$_3^{1.63-}$ and then  
  the valence is reduced significantly  from the formal valence. 
 \cite{Cohen}
 Such a prediction  is  consistent with the XPS experiment.
\cite{Hudson}
 Further 
 the effect of correlation is also claimed by  
 the fact that 
 dynamical effective charges  for oxygen and titanium 
  are large due to the  covalence.
\cite{Zhong,Resta1}

In terms  of exact  diagonalization, 
 the effect of the electron correlation has been examined 
    for  a one-dimensional half-filled Hubbard model  
  with  alternating potential, which is  caused  by   two kinds of atoms 
 and   induces the difference of the site energy.
\cite{Egami,Ishihara} 
 When the repulsive interaction increases or the level separation 
   decreases,  the band  (ionic-covalent) insulator 
   undergoes a  transition 
  into  the Mott insulator  followed by a jump of  electron numbers  
    of  each  lattice sites.   
  In addition to 
   the electron-lattice interaction,
  the electronic polarizability is strongly enhanced 
   near the phase boundary due to the correlation.
\cite{Ishihara2}

 Polarization induced by a sublattice displacement,   
  has been examined for  a one-dimensional two-band Hubbard model 
 at half-filling.
\cite{Resta1,Resta2}   
 The number of parameters is   reduced by setting 
 equal  values of  the  Hubbard $U$  on 
   two kinds of atoms of  oxygen and generic cation,
 while  the site energies are different. 
  At a critical value of interaction corresponding to 
   a transition from  a band insulator to  a Mott insulator,
 the static ionic charge  is continuous 
 but the polarization and the dynamic charge are discontinuous showing 
 the Berry's phase
\cite{Resta_review}
 associated with the macroscopic polarization 
   as the primary order parameter.    
 The discontinuity disappears by  removing the  finite-size effect, 
  i.e., displacing the coarse mesh in order to 
  avoid the  $k=0$ singular point and  taking the large number 
     of $k$-points 
 where $k$ is the quasicell momenta. 

  The model which is the same as 
 ref.~\citen{Resta1} but has the infinite length,   
    has been explored by use of mean-field approximation    
   in order to clarify the nature of electronic states and 
          the relation of the Berry's phase to the phase transition.
\cite{Ortiz}  
 The ground state has 
  antiferromagnetic ordering for large repulsive  interaction  
  indicating a magnetic instability  
    while  the state is nonmagnetic for small interaction.
  There exists the gap to the charge excitations independently of 
    the value of repulsive interaction while  
     the system  is replaced by antiferromagnet 
   with gapless spin excitations    in strong coupling regime  
  and has a gap  to all excitations 
   in the opposite limit. 
  Although the magnetic state with the  gapless spin  excitation 
 is known  for the conventional Hubbard model with the repulsive 
 interaction, it is not clear if 
 such an excitation still exists even for the small limit of 
  the alternating potential.   
 Therefore  
 the method  beyond the mean-field theory is needed to study 
  the model with  the infinite length.

 In order to   treat  strictly 
 one-dimensional fluctuations and  electron correlation,
  we apply, in the present paper,   a method of  
  the bosonization to the system, which is the same as 
 refs.~\citen{Resta1} and \citen{Ortiz}.   
 Further, the renormalization group method is utilized where 
  the electronic states of the band insulator and   the Mott insulator 
 are  determined by    the relevance and  the irrelevance of  
  alternating potential.
 In \S 2, formulation for deriving the renormalization group equations 
  is given in terms of the phase variable, 
  which is based on the bosonization. 
In \S 3, the transition from the band insulator to the Mott insulator 
  is calculated as the function of 
 on-site electron-electron  interaction and 
 the energy difference between two-sites. 
 In \S 4, summary is given and  
 the effect of the alternation of  the electron hopping energy 
 given by the Su-Schrieffer-Heeger model,\cite{Su} is briefly discussed.

\section{Formulation}

 The Hamiltonian of one-dimensional half-filled  Hubbard model 
  with the  alternating potential is given by 
\begin{eqnarray}
{\cal  H} &=& - t \, \sum_{j, \sigma} 
   \left(  c_{j, \sigma}^{\dagger} 
    \, c_{j+1, \sigma} + \mbox{\rm h.c.} \right)
+ W_0 \sum_{j,\sigma} \,(-1)^j \, n_{j,\sigma}
+ U \sum_{j} \, n_{j, \uparrow} \, n_{j, \downarrow}
\virg
\label{eq:H}
\end{eqnarray}
 where 
   $c_{j,\sigma}^\dagger$  
    denotes a creation operator for the electron at the $j$-th site
  with spin $\sigma (=\uparrow, \downarrow) $.
  $n_{j,\sigma} = c_{j,\sigma}^\dagger c_{j,\sigma}$.
 Quantities $t$, $W_0$ and $U$   are energies for the transfer integral,  
 the alternating potential and  the 
 on-site electron-electron Coulomb repulsive interaction, respectively. 

 First, eq.~(\ref{eq:H}) is rewritten by use of 
   the Fourier transform, 
 $c_{k, \sigma}= 1/\sqrt{N} \,
             \sum_{j} \, \e^{-\i k R_j} \, c_{j,\sigma}$,
   where $R_j$ is the location of the $j$-th lattice site and 
  $N$ is the total number of the lattice site.
 The first and second terms  of  eq.~(\ref{eq:H})  are  expressed as,
$\sum_{k, \sigma} \varepsilon_k \,
     c_{k, \sigma}^{\dagger}  \, c_{k, \sigma}
+ W_0  \sum_{k>0,\sigma}
       \left(
             c_{k,\sigma}^\dagger c_{k-\pi/a,\sigma} 
           + c_{k-\pi/a,\sigma}^\dagger c_{k,\sigma} 
       \right)$
  where $\varepsilon_k = -2t \cos ka$ and $a$ is the lattice constant.
 The dispersion of the kinetic energy, $\varepsilon_k$,  
  around the Fermi momentum  is linearized as  $\vf(pk - \kf)$,
  where $\vf(=2 ta)$ and $\kf(=\pi/2a)$ denote the Fermi velocity 
  and the Fermi momentum, respectively.
  The fermion operator around the Fermi momentum is rewritten as 
  $c_{k,p,\sigma}(=c_{k+p\kf,\sigma})$, where 
    $p (=+$ and $-)$ denotes the branch  
  for the right moving and left moving  electrons. 
The third term of eq.~(\ref{eq:H}) is rewritten  as ($L=Na$)
\begin{eqnarray}
{\cal H}_{\rm int} &=&   
\frac{g_{1}}{L}  
\sum_{k_1, k_2,q, p}
    c_{k_1,p,\uparrow}^\dagger \, c_{k_2,-p,\downarrow}^\dagger \, 
    c_{k_2+q,p,\downarrow} \, c_{k_1-q,-p,\uparrow}
\nonumber \\
&+& 
\frac{g_{2}}{L}  
\sum_{k_1, k_2,q, p}
    c_{k_1,p,\uparrow}^\dagger \, c_{k_2,-p,\downarrow}^\dagger \, 
    c_{k_2+q,-p,\downarrow} \, c_{k_1-q,p,\uparrow}
\nonumber \\
&+& 
\frac{g_{3}}{L}  
\sum_{k_1, k_2,q, p}
    c_{k_1,p,\uparrow}^\dagger \, c_{k_2,p,\downarrow}^\dagger \, 
    c_{k_2+q,-p,\downarrow} \, c_{k_1-q, -p,\uparrow}
\nonumber \\
&+& 
\frac{g_{4}}{L}  
\sum_{k_1, k_2,q, p}
    c_{k_1,p,\uparrow}^\dagger \, c_{k_2,p,\downarrow}^\dagger \, 
    c_{k_2+q,p,\downarrow} \, c_{k_1-q,p,\uparrow} 
\virg
\label{eq:Hint}
\end{eqnarray}
where
$g_{1} = g_{2} = g_{3}= g_{4}= Ua $ and
 these quantities denote 
 coupling constants for the backward scattering ($g_1$), 
  forward scattering between the opposite branch ($g_2$), 
 umklapp scattering ($g_3$) and 
  forward scattering within the same branch ($g_4$),  
 respectively.

Next the bosonization method
\cite{Solyom}  
 is applied to eq.~(\ref{eq:H}).
 The   phase variables, 
  $\theta_{\rho \pm}$ and  $\theta_{\sigma \pm}$,
    expressing     fluctuations  of the charge density and spin density
   are given by  
\begin{eqnarray}
                           \label{eq:theta_rho}
\theta_{\rho \pm} (x)  
         &=& \frac{1}{\sqrt{2}} \sum_{q\neq 0} \frac{\pi \i}{qL}
             \e ^{-\frac{\alpha}{2}|q|-\i qx} 
\sum_{k,\sigma}
    \left(  c_{k+q,+,\sigma}^\dagger c_{k,+,\sigma}
        \pm c_{k+q,-,\sigma}^\dagger c_{k,-,\sigma}
    \right)
\virg \\
\theta_{\sigma \pm} (x)
         &=& \frac{1}{\sqrt{2}} \sum_{q\neq 0} \frac{\pi \i}{qL}
             \e ^{-\frac{\alpha}{2}|q|-\i qx}
 \sum_{k,\sigma} \sigma
    \left(  c_{k+q,+,\sigma}^\dagger c_{k,+,\sigma}
        \pm c_{k+q,-,\sigma}^\dagger c_{k,-,\sigma}
    \right)
        \label{eq:theta_sigma}
\virg
\end{eqnarray}
  where  $\sigma = +$($-$) corresponds to $\uparrow$($\downarrow$) 
and 
$  [\theta_{\nu +}(x),\theta_{\nu' -}(x')]
 = \i \pi \delta_{\nu, \nu'}\,{\rm sgn}(x-x')$. 
By using these phase variables, we rewrite the field operator,
$\psi_{p,\sigma}(x)(= L^{-1}  \sum_k \e^{\i p\kf x+\i kx} 
   c_{k,p,\sigma})$,
    as 
\begin{eqnarray}
\psi_{p,\sigma}
&=& 
 \frac{1}{\sqrt{2\pi \alpha} }
\exp \left( \i pk_{{\rm F}}x 
 + \i \Theta _{p,\sigma} \right) 
   \exp \left( \i \pi \Xi_{p,\sigma} \right) \virg
\label{eqn:field} \\ 
\Theta _{p,\sigma}
 & = &  
\frac{1}{2}
  [
     p \, \theta_{\rho +} + \theta_{\rho -} 
   + \sigma ( p \, \theta_{\sigma +} + \theta_{\sigma -} )
                      ]\virg
              \label{eq:thta}
\end{eqnarray}
 where
   $\alpha$ is of the order of the lattice constant.  
 The phase factor 
  in eq.~(\ref{eqn:field}), 
  is chosen  as 
$\Xi_{p,+} = p(\N_{+,+} + \N_{-,+})/2$ and
$\Xi_{p,-} = (\N_{+,+} + \N_{-,+}) 
              + p (\N_{+,-} + \N_{-,-})/2$
with  the number operator,  $\N_{p,\sigma}$, 
 in order to satisfy the anticommutation relation. 
 Such a choice of   $\Xi_{p,\sigma}$ 
 conserves a sign of the 
  nonlinear terms, which is obtained from the interaction terms 
 of eq.~(\ref{eq:Hint}).
Based on these bosonic fields, 
  eq.~(\ref{eq:H}) is rewritten as \cite{Suzumura_PTP}
\begin{eqnarray} 
{\cal H} &=& 
  \frac{v_\rho}{4\pi} \int \hspace{-1mm} \dx
 \Bigl[
   \frac{1}{K_\rho} \left(\partial \theta_{\rho +} \right)^2
         +  K_\rho  \left(\partial \theta_{\rho -} \right)^2
 \Bigr]
+
    \frac{g_{3}}{2\pi^2 \alpha^2}  \int \hspace{-1mm}\dx
    \cos 2\theta_{\rho +} 
\nonumber \\ 
&+&
  \frac{v_\sigma}{4\pi} \int \hspace{-1mm} \dx
 \Bigl[
   \frac{1}{K_\sigma} \left(\partial \theta_{\sigma +} \right)^2
         +  K_\sigma  \left(\partial \theta_{\sigma -} \right)^2
 \Bigr]
+
    \frac{g_{1}}{2\pi^2 \alpha^2}  \int \hspace{-1mm}\dx
    \cos 2\theta_{\sigma +} 
\nonumber \\ 
&+&
  \frac{2W_0}{\pi \alpha}  \int \hspace{-1mm}\dx
    \cos \theta_{\rho +} 
    \cos \theta_{\sigma +} 
\virg
\label{eq:Hphase}
\end{eqnarray}
 where  
$v_\rho   = [(2\pi\vf+g_4)^2-g_2^2]^{1/2}/2\pi$,
$K_\rho   = [(2\pi\vf+g_4-g_2)/(2\pi\vf+g_4+g_2)]^{1/2}$,
$v_\sigma = [(2\pi\vf-g_4)^2-g_1^2]^{1/2}/2\pi$ and
$K_\sigma   =  [(2\pi\vf-g_4+g_1)/(2\pi\vf-g_4-g_1)]^{1/2}$.

 Now we derive renormalization group equations by examining 
  response functions, 
$
R_{A}(x_1-x_2,\tau _1-\tau _2) 
\equiv 
\langle T_\tau \O_A(x_1,\tau_1) \, \O_A^\dagger (x_2,\tau_2) \rangle 
$,
 where   
  $\tau_j$ is the imaginary time and 
   $\O_A$ denotes  several kinds of  order parameters.
 Response functions    
 are evaluated perturbatively by expanding  
 nonlinear terms of eq.~(\ref{eq:Hphase}).
 From the assumption  that response functions are  
    invariant for   
   $\alpha \to \alpha '=\alpha \e^{\d l}$,
\cite{Giamarchi_JPF,Giamarchi_PRB}   
 renormalization group equations 
  for  
$K_{\rho}, K_{\sigma}, g_3, g_1$ and $W_0$  
  are derived as  (Appendix A) 
\begin{eqnarray}
\dl K_\rho (l)&=& 
- 2 \, G_3^2(l) \, K_\rho^2(l) 
-  G_{W}^2(l) \, K_\rho^2(l) 
\virg
\label{eq:Krho}
\\
\dl K_\sigma (l) &=& 
- 2 \, G_{1}^2(l) \, K_\sigma^2 (l) 
- G_{W}^2(l) \, K_\sigma^2(l) 
\virg
\label{eq:Ksigma}
\\
\dl G_{3}(l) &=& 
\bigl[2-2K_\rho(l) \bigr]\, G_{3}(l) - G_{W}^2(l)
\virg
\label{eq:G3}
\\
\dl G_{1} (l)&=& 
\bigl[2-2K_\sigma(l) \bigr]\, G_{1}(l) - G_{W}^2(l)
\virg
\label{eq:G1}
\\
\dl G_{W}(l) &=& 
\bigl[2-K_\rho(l)/2 -K_\sigma (l)/2 \bigr]\, G_{W}(l)
\nonumber \\ &&
 - G_{3}(l) \,\, G_{W}(l)  - G_{1}(l) \,\, G_{W}(l) 
\label{eq:GW}
\virg
\end{eqnarray}
 where initial conditions are chosen as   
  $K_\nu (0) = K_\nu$ ($\nu= \rho$ and $\sigma$),
  $G_{3} (0) = g_{3}/2\pi\vf$,
  $G_{1} (0) = g_{1}/2\pi \vf$ 
  and $G_{W} (0) = W_0/(\vf\alpha^{-1})$.
 For the simplicity, 
  we replaced $v_\rho$ and $v_\sigma$ by $\vf$,
 in  eqs.~(\ref{eq:Krho})-(\ref{eq:GW}) 
 since   the  deviation of the numerical factor is small.

The calculation of $K_{\sigma}(l)$ in eq.~(\ref{eq:Ksigma}) is 
  performed by the following procedure.
When we substitute $K_{\sigma}(l)=1+G_1(l)$ for r.h.s. of 
  eqs.~(\ref{eq:Ksigma}) and (\ref{eq:G1}) and retain the 
  coupling constants up to the second order,
    eq.~(\ref{eq:Ksigma}) becomes equal to eq.~(\ref{eq:G1}). 
In this case, the fixed point of $K_{\sigma}(l)$  
   with $W_{0}=0$ is given by  unity 
   showing the validity of  the SU(2) symmetry, i.e,  
     $K_{\sigma}(\infty) = 1$.  
Therefore eqs.~(\ref{eq:Ksigma}) and (\ref{eq:G1}) are
  treated by such an expanded form for $K_{\sigma}(l) > 1$
  while we use the original form of eq.~(\ref{eq:Ksigma}) for 
  $K_{\sigma}(l) < 1$.

 We note the electronic states obtained from 
     eq.~(\ref{eq:GW}). 
In the absence of the $U$-term,  the state of 
  eq.~(\ref{eq:H}) has   a gap, $2 W_0$, at $k = \pm \kf$ of the band
  while that of eq.~(\ref{eq:H}) with $W_0=0$ exhibits a 
   charge gap, $\Delta_{\rho}$, due to half-filling 
    corresponding to a Mott gap.  
 Both of these cases  are insulating states. 
  The former  is the  band insulator with 
    the relevant  $G_{W}(l)$, which 
     increases infinity for large $l$. 
 The latter is the Mott insulator with the relevant $G_3(l)$. 
  In the presence of both  $U$-term and  $W_0$-term, 
   there are  two cases of    
        the relevant $G_{W}(l)$ and  the irrelevant $G_{W}(l)$. 
 We use the terminology of the band insulator for 
  the relevant $G_{W}(l)$ and 
 the Mott insulator for the irrelevant $G_{W}(l)$ 
 although the latter case  denotes actually a crossover to the Mott 
 state, i.e, the spin gap decreases rapidly to zero as shown 
  in the next section.

Here we examine the order parameters, which are characteristic 
  of these insulating states. 
 Since  
  the phase of $\theta_{\rho +}$ is locked at $\pi$ or 0
 $ (\pm \pi/2)$   
 for the relevant (irrelevant) $G_{W}(l)$,  
 we can consider the order parameter of the band insulator
 (the Mott insulator) 
  as  $\langle \cos \theta_{\rho+} \rangle$
 ($ \langle \sin \theta_{\rho+} \rangle$).
 The finite value of  $\langle \cos \theta_{\rho+} \rangle$ 
  denotes the ionic state 
 since the $W_0$-term of eqs.~(\ref{eq:H}) and (\ref{eq:Hphase})  
  is proportional to 
 the difference between  electron numbers of  two kinds of 
 lattice sites. 
  As for the locking of  $\theta_{\sigma +}$, 
  the effect of the $W_0$-term is much stronger than that of 
  the $g_1$-term in eq.~(\ref{eq:Hphase}),
   because the $g_1$-term becomes irrelevant for 
    $W_0=0$. 
 Thus we always have the finite value of 
   $\langle \cos \theta_{\sigma+} \rangle$.
 We note that the quantity 
   $\lan \cos \theta_{\rho +}\ran \lan\cos \theta_{\sigma +} \ran$
    is negative due to the positive value 
   of $W_0$ in eq.~(\ref{eq:Hphase}) and that 
   there is a degeneracy of a state with  
  the positive $\lan \cos \theta_{\rho +}   \ran $ and 
  the negative $\lan \cos \theta_{\sigma +} \ran $
 and a state with 
 the negative $\lan \cos \theta_{\rho +}   \ran $ and  
 the positive $\lan \cos \theta_{\sigma +} \ran $. 
For studying these properties, 
 we calculate  response functions, 
  $R_{\cos \theta_{\nu +}}(\br_1-\br_2) 
    =
    \langle T_\tau 
     \cos \theta_{\nu+}(\br_1)  
     \cos \theta_{\nu+}(\br_2) 
    \rangle $ ($\nu=\rho$ and $\sigma$). 
By use of the renormalization group technique, 
  these  response functions are derived as (Appendix A)
\begin{eqnarray}
R_{\cos \theta_{\rho +}}(r) 
&=&
\exp \left[- \int_0^{\ln(r/\alpha)} \d l \,\,
  \bigl(K_\rho (l) +2G_3(l) \bigr)\right] \virg
\label{eq:response_r}
\\
R_{\cos \theta_{\sigma +}}(r) 
       &=&
\exp \left[- \int_0^{\ln(r/\alpha)} \d l \,\,
  \bigl(K_\sigma (l) +2G_1(l) \bigr)\right] \virg 
\label{eq:response_s}
\end{eqnarray}
where $r =[x^2+(\vf \tau)^2]^{1/2}$.
For the band insulator the limiting values of both 
  $R_{\cos \theta_{\rho +}}(r)$ and $R_{\cos \theta_{\sigma +}}(r)$
  are finite,
  while 
  only that of  $R_{\cos \theta_{\sigma +}}(r)$
  becomes   finite for the Mott insulator.

\section{Band insulator vs. Mott insulator}

 For the calculation of characteristic energy in terms of  
  renormalization group  equations, 
  it is crucial to determine the magnitude of 
     the cutoff parameter, $\alpha$, which is chosen  conventionally 
      as   $\alpha = a/\pi$.\cite{Solyom} 
 We estimate  $\alpha$ from  the  response function for  charge density wave 
 with a momentum close to 2$\kf$  
  as follows. 
 Such a response function with $W_0 = U = 0$ can be calculated 
    explicitly 
  by both the conventional method and the bosonization method 
 (Appendix B). 
 From  the comparison of these two quantities for the half-filled band, 
 we obtain $\alpha \simeq  \pi/(1.273 a)$.
  Thus   characteristic energy corresponding to 
     $l$ is given by   $\vf\alpha^{-1} \exp [-l]$.

\begin{fullfigure}
\hspace*{4cm}
\epsfxsize=7.5cm
\epsffile{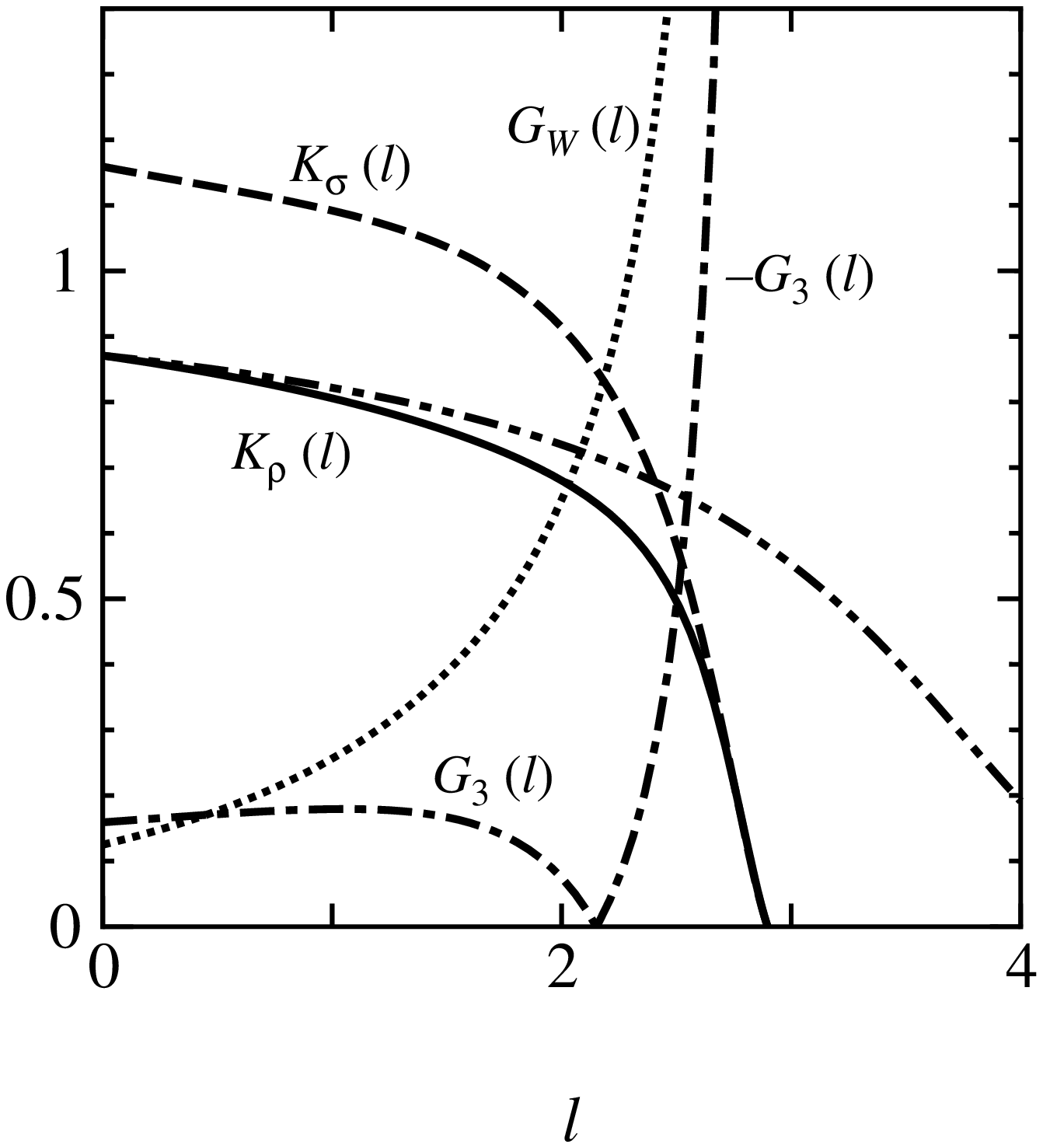}
\vspace*{-1cm}
\caption{
The $l$-dependence of coupling constants, $K_\rho (l)$, $K_\sigma (l)$,
  $G_3(l)$ and $G_W(l)$,
  with fixed  $U/t= 2$ and $W_0/t=1$.
 The 2-dot-dashed curve denotes $K_\rho (l)$
 for  $W_0/t=0$.
}
\vspace*{2cm}
\hspace*{4cm}
\epsfxsize=7.5cm
\epsffile{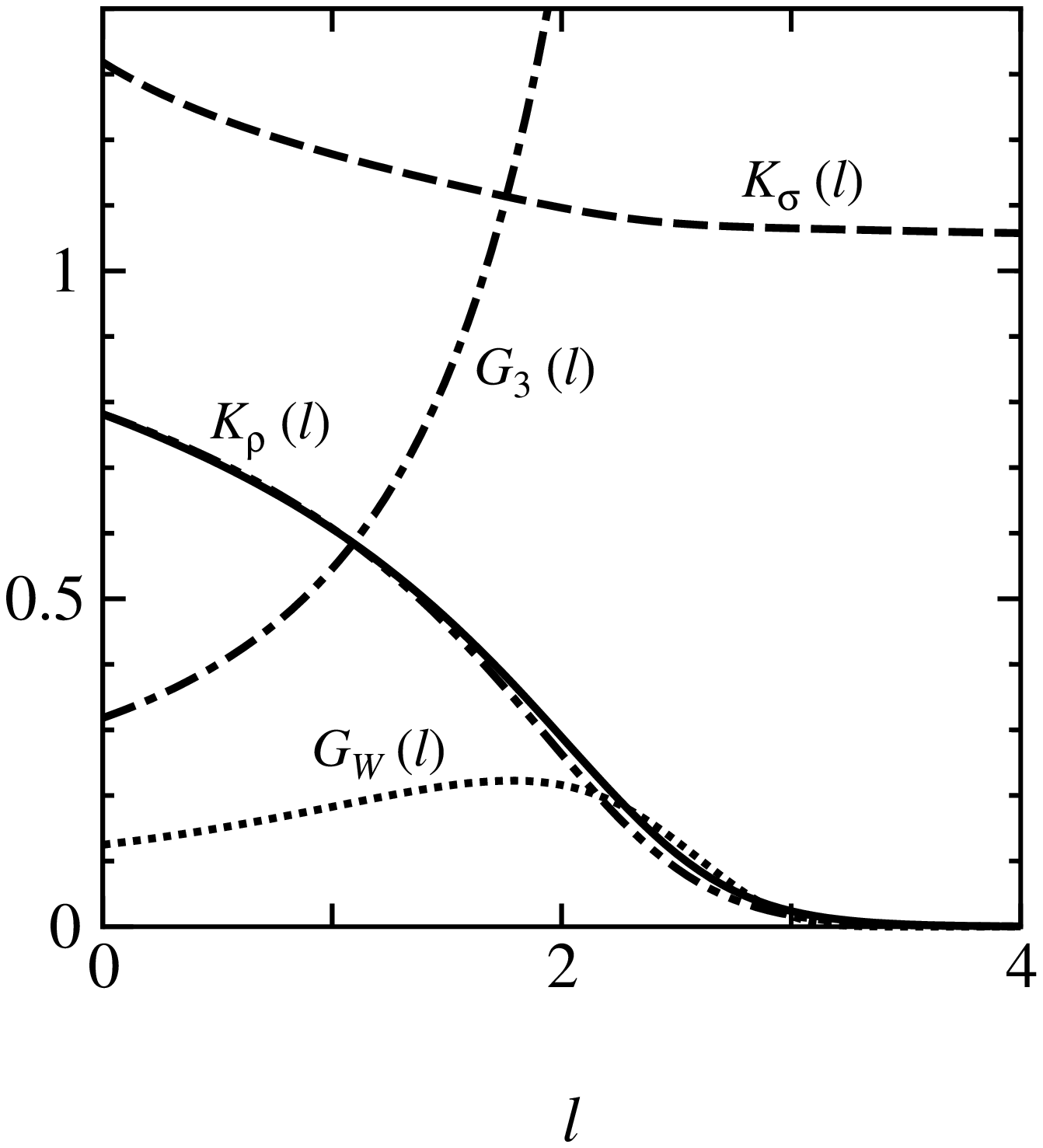}
\vspace*{-1cm}
\caption{
The $l$-dependence of coupling constants, $K_\rho (l)$, $K_\sigma (l)$,
  $G_3(l)$ and $G_W(l)$,
   with fixed  $U/t= 4$ and $W_0/t=1$ 
 where the notations are the same as Fig.~1. 
}
\end{fullfigure}
 In Fig.~1, quantities $K_{\rho}(l)$(solid curve), 
 $K_{\sigma}(l)$(dashed curve), $G_{3}(l)$(dot-dashed curve) and 
 $G_{W}(l)$(dotted curve) 
 are shown as a function of $l$  for $U/t = 2$ and $W_0/t=1$.  
 With increasing $l$, $G_W(l)$ increases monotonically 
  and then becomes relevant leading to the band insulator. 
The change of sign of $G_3(l)$ occurs at $l \simeq  2.1$ due to the 
  relevant $G_W(l)$.
The positive  $G_3(l)$  originates in the umklapp scattering, $g_3$, 
    and the negative $G_3(l)$ comes from the higher harmonics  
    induced by the $W_0$-term.    
With increasing $l$, 
   $K_{\rho}$ decreases monotonically since both 
    $G_{3}(l)$ and $G_{W}(l)$ reduce  $K_{\rho}(l)$. 
The reduction of $K_\rho(l)$ by $G_W(l)$ is understood from the fact 
  that $K_{\rho}(l)$  is suppressed noticeably 
    compared with the case of $W_0 = 0$ (2-dot-dashed curve).  
The  $l$-dependence  of $G_1(l)$ is similar to 
   that of  $G_3(l)$, i.e.,  
   the positive $G_{1}(l)$ changes the sign with increasing $l$. 
With increasing $l$,
   $K_{\rho}(l)$ and $K_{\sigma}(l)$ merge  each other and 
  their magnitudes are reduced  to zero. 
This behavior indicates the formation of both charge gap, $E_{g}$, 
  and  spin gap, $\Delta_{\sigma}$, which originate in  the $W_0$-term.

In Fig.~2, the numerical result for $U/t = 4$ and $W_0/t=1$ is shown
  where the notations are the same as Fig.~1.
The behavior of $K_\sigma(l)$, $G_3(l)$ and $G_W(l)$ is 
  quite different compared with that of Fig.~1 
  although that of $K_\rho (l)$ is qualitatively the same.
The difference between $K_{\rho}(l)$ (solid curve) and 
  that of $W_{0}=0$ (2-dot-dashed curve) is very small 
  due to the irrelevant $G_W(l)$. 
 The effect of 
 the $W_0$-term reducing the charge gap is seen from the fact that 
    $K_{\rho} (l)$ in the presence of 
  $W_0$ (solid curve) is slightly larger than 
   that of $W_0=0$ (2-dot-dashed curve). 
With increasing $l$, $G_{W}(l)$ takes a maximum and decreases to zero.
Thus the effect of  $W_{0}$ becomes negligibly small 
     for the large $l$ (i.e., small energy).  
Such an irrelevant $G_{W}(l)$ leads to a state similar to 
   the Mott insulator induced by the umklapp scattering. 
The quantity $G_3(l)$ is always positive and increases monotonically
  from the initial value $g_3$.
The $l$-dependence of $K_{\sigma}(l)$ is 
  similar to that with $W_{0}=0$, i.e., $K_\sigma(l) \gsim 1$
  in the range of Fig.~2.
However a small spin gap does exist since $K_{\sigma}(l)$ reduces to 
  zero for the large $l$.

\begin{fullfigure}
\hspace*{4cm}
\epsfxsize=7.5cm
\epsffile{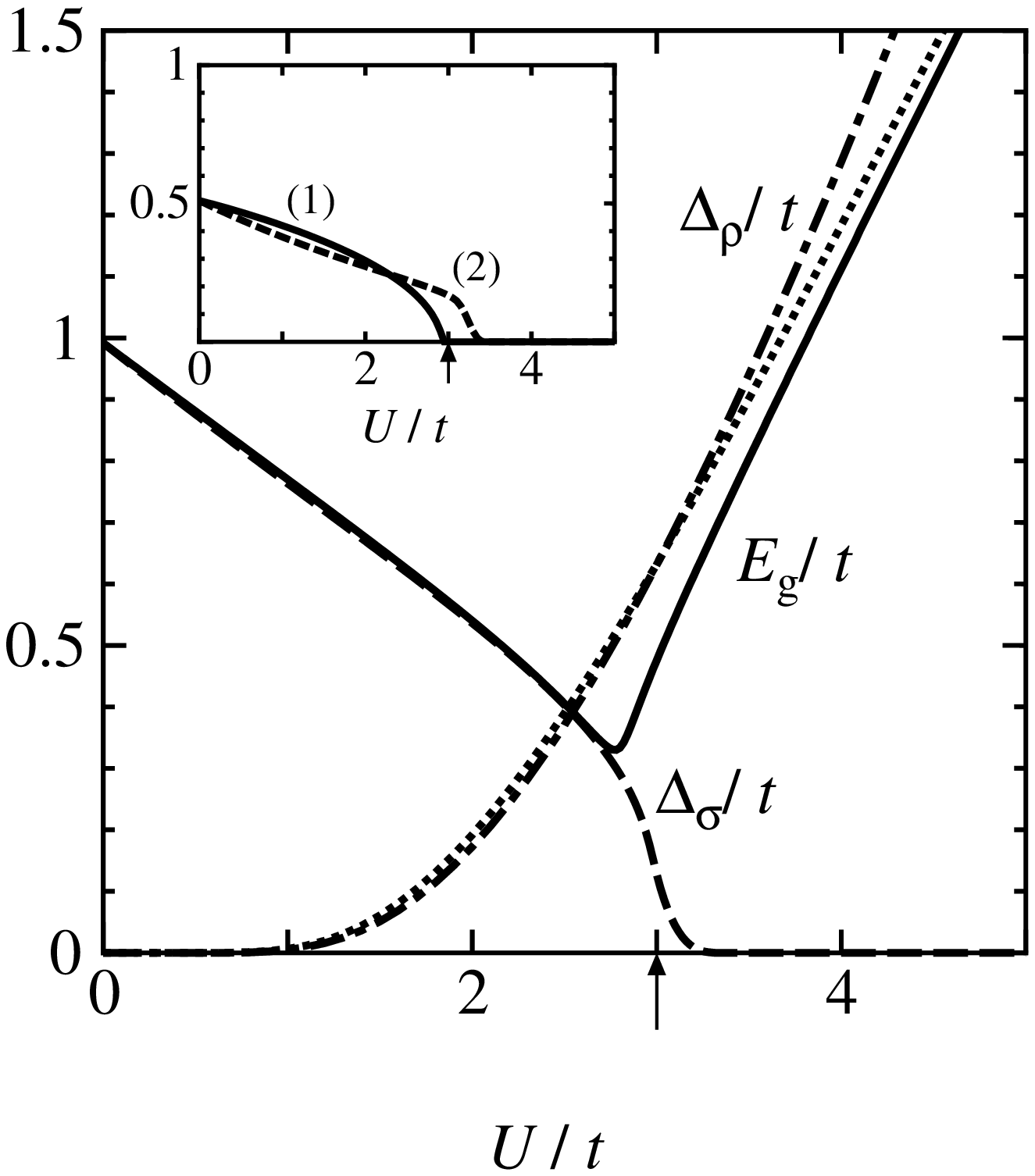}
\vspace*{-1cm}
\caption{
The $U$-dependence of $E_g/t$ and $\Delta_\sigma/t$
  is shown with fixed $W_0/t=1$. 
The arrow denotes a location for a critical value,
  $U_c/t (\simeq 3.0)$, where     
   the  band insulator is obtained for $U< U_c$ and  
    the Mott insulator is obtained for $U>U_c$.
The charge gap, $\Delta_\rho/t$, 
  for $W_0/t=0$ is shown by the dotted curve (the present calculation)
  and the dot-dashed curve (the exact solution)\cite{Lieb}.
In the inset, the $U$-dependence of 
 $|\lan \cos \theta_{\rho +} \ran|$ and
  $|\lan \cos \theta_{\sigma +} \ran|$ 
 is shown by curve (1) and  curve (2)  respectively where 
 the arrow corresponds to  $U_c$.
}
\end{fullfigure}
Based on the calculation given by Figs.~1 and 2, 
  it turns out that there is a critical value of $U = U_c$, 
   where  the band insulator is obtained for $U < U_c$ 
  and the Mott insulator is obtained for $U > U_c$.  
 In Fig.~3, the $U$-dependence of excitation gap and 
   $|\lan \cos \theta_{\nu +} \ran|$  is shown. 
  The  charge gap ($E_{g}$) and  spin gap ($\Delta_{\sigma}$) 
  are evaluated  by the formula, 
    $E_g = \vf \alpha^{-1} \exp [-l_{\rho}]$ and 
      $\Delta_\sigma = \vf \alpha^{-1} \exp [-l_{\sigma}]$  where 
   $l_{\nu}$ ($\nu=\rho$ and $\sigma$) is chosen as  
    $K_\nu(l_{\nu}) = 0.3$. 
Such a choice is reasonable since, in the absence of $W_0=0$, 
 the charge gap of the present calculation 
 (dotted curve)  coincides well 
  with the exact one (dot-dashed curve).\cite{Lieb}
 The $U$-dependence of $E_g$ and $\Delta_{\sigma}$ 
  is shown by solid curve and dashed curve, respectively.  
The arrow denotes the location for a critical value of 
  $U_{c}/t \simeq 3.0$, 
  which separates the band insulator from the Mott insulator.
There is a  minimum of $E_{g}$ (solid curve) 
    where the corresponding $U$ is slightly smaller than $U_c$. 
 Such a minimum  arises from  
  a competition  between the $W_0$-term and  the $g_3$-term. 
 For large $U$, the magnitude of $E_{g}$ moves close  to 
  $\Delta_{\rho}$(dotted curve), 
  which denotes the  charge gap in the absence of $W_{0}$.    
 With increasing $U$, the spin gap $\Delta_{\sigma}$ decreases but 
  takes a finite value even for $U > U_c$. 
 For  small $U$, 
  one finds $E_g \simeq \Delta_{\sigma}$ indicating that 
 the difference between the  magnitude of the charge gap and 
 that of the spin gap is negligibly small.   
Thus it turns out that 
  the behavior  of the  charge gap  is quite different from 
  that of spin gap.

In the inset of Fig.~3, 
   the $U$-dependence of $|\lan \cos \theta_{\rho +}\ran|$ and  
  $|\lan \cos \theta_{\sigma +} \ran|$ 
 is shown 
  by solid curve (1) and   dashed curve (2), 
   respectively where 
   the arrow denotes $U=U_c$. 
 The quantity $\lan \cos \theta_{\nu +}\ran$ 
  ($\nu = \rho$ and $\sigma$),
     is defined by 
 $ \lan T_\tau \cos \theta_{\nu +}(x) \cos \theta_{\nu +}(0) \ran
  \rightarrow  \lan \cos \theta_{\nu +} \ran^2 $ 
 for $|x| \rightarrow \infty$.
In the context of the present calculation,  such a value is calculated 
 from the minimum value  of response function 
 due to the second order renormalization group equation.  
With increasing $U/t$,  
   $|\lan \cos \theta_{\rho +} \ran|$ decreases and is reduced to zero 
  at $U=U_c$,  corresponding to   
   the transition  from  the band insulator
    to the Mott insulator, 
   where the locking of phase for the former (latter) state  
 is given by $ \theta_{\rho +} =\pi$ or $0$ 
    ($\theta_{\rho +}=\pm \pi/2$). 
The finite value of  $\lan \cos \theta_{\rho +} \ran$ 
  corresponds  to the relevance of $G_{W}(l)$, i.e, 
  the band insulator. 
 In contrast to 
  $\lan \cos \theta_{\rho +} \ran$, 
  we find that 
 $\lan \sin \theta_{\rho +} \ran$ becomes finite 
 for the irrelevance of $G_{W}(l)$, i.e.,   $U > U_c$
 due to the change of the locked  
  $\theta_{\rho +}$ from 0 to $\pm \pi/2$. 
 The quantity $|\lan \cos \theta_{\sigma +} \ran|$ decreases 
  but remains finite even for $U > U_c$ 
  showing the similarity to $\Delta_{\sigma}$. 
We note that
    $|\lan \cos \theta_{\rho +} \ran| \propto \sqrt{|W_0|} $ 
  for  $U = 0$ and  
          small $|W_0|/t \lsim 2   $. 
 Such a result is compatible with the fact that 
  the conventional band gap with  $U = 0$, 
    is proportional  to $W_0$. Then 
  the present calculation is valid  for  $W_0/t\lsim 2 $. 
  The $U$-dependence of $E_g$ and  
   $\lan \cos \theta_{\rho +} \ran$ 
 ($\Delta_{\sigma}$ and  $\lan \cos \theta_{\sigma +} \ran$)
 shows that 
    a transition (a crossover)  from  the band insulator 
     to the Mott insulator appears due to   
       the  competition between the 
   alternating potential and the electron-electron interaction.

\begin{fullfigure}
\hspace*{4cm}
\epsfxsize=8cm
\epsffile{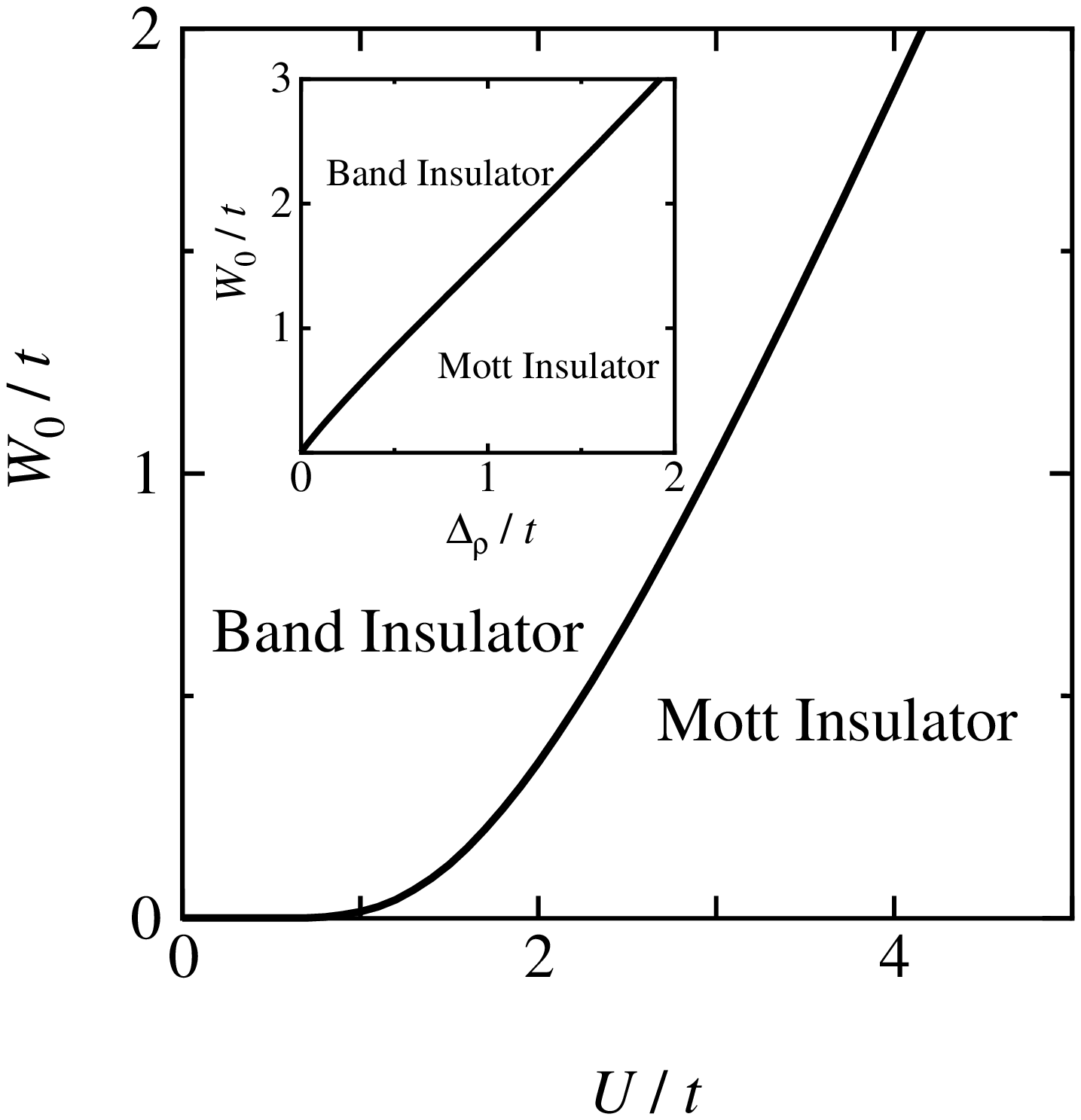}
\vspace*{-1cm}
\caption{
Phase diagram for the band insulator 
  and the Mott insulator on the plane of $U/t$ and $W_0/t$.
 The inset shows  the same phase diagram on the plane of 
  $\Delta_\rho/t$ and $W_0/t$.
}
\end{fullfigure}
 From the calculation of $U_{c}$ similar to Fig.~3 
     with several choices of  $W_0$,
   a phase diagram of the band insulator and Mott insulator 
  is shown  on the plane of  $U/t$ and $W_0/t$  in Fig.~4. 
 Since the the boundary between these two states is convex downward, 
   the effect of $W_0$ is much larger than $U$ for small $U$. 
  The  boundary, 
      which is well reproduced by a formula,  $\exp[-2\pi t/U]$, 
       for small $U/t$, 
 indicates a fact that the characteristic energy 
   for the competition  is not $U$ but 
  the charge gap $\Delta_{\rho}$.
   In the inset, the same   phase diagram is shown   
    on the plane of $\Delta_{\rho}/t$ and $W_0/t$.  
 From the result that  
  $W_0/\Delta_\rho = 1.6 \sim 1.7$ for $0.3<\Delta_\rho/t <2.0$, 
   it is found that  the boundary between the band insulator and 
    Mott insulator is determined by 
        the competition between    $\Delta_{\rho}$ and $W_0$.

\section{Summary and Discussion}

In the present paper, 
  we have examined the effect of the alternating potential, $W_0$,  
  on a one-dimensional half-filled  
    Hubbard model with the repulsive interaction, $U$.
  The property of the model with  the infinite length is calculated 
    by use of the methods of  bosonization and renormalization group.  
  With increasing $U$, a second order phase transition 
   followed by the continuous change of the order parameters 
    occurs from the  band insulator into 
      the Mott insulator at $U_c$  
       where the former (latter)  corresponds to 
        the relevant (irrelevant) $G_{W}(l)$.  
 Such a transition is demonstrated  from the fact that  
    $\lan \cos \theta_{\rho +} \ran$ 
 becomes zero 
   for  $U \rightarrow  U_c -0 $. 
      The band insulator denotes the ionic state since  
    $\lan \cos \theta_{\rho +} \ran \not= 0$ 
  in addition to  $\lan \cos \theta_{\sigma +} \ran \not= 0$. 
  Phases of  the band insulator are locked at  
     $\theta_{\rho +}= \pi$ (or 0) and  $\theta_{\sigma +}=0$ (or $\pi$) 
   respectively. 
 For the Mott insulator, phases are locked at 
    $\theta_{\rho +}= \pi/2$ or $-\pi/2$    
      and  $\theta_{\sigma +}=0$ or $\pi$ due to the presence of 
 the spin gap even if 
  $G_{W}(l)$ is irrelevant. 
 The charge gap, $E_g$, takes a minimum  
 and the  spin gap $\Delta_{\sigma}$  exhibits a crossover 
 around the transition point of $U=U_c$. 
 The phase diagram  of  band insulator and  Mott insulator   
   has been obtained on the plane of $U$ and $W_0$.

 Here we compare  the boundary calculated in the present paper 
  with   those   obtained by the numerical diagonalization. 
 In ref.~\citen{Resta1}, in which the same  model is treated, 
   the boundary between  band insulator and Mott insulator 
     is calculated as $U_c/t \simeq 2.27$  for $W_0/t \simeq 0.57$ 
  while our result shows $U_c/t \simeq 2.4$ for $W_0/t = 0.57$. 
 In refs.~\citen{Egami} and \citen{Ishihara} 
  which introduced two kinds on-site potentials 
    $U_B < U_A$  in addition to the present model, 
 they obtained  $U_{Bc}/t \simeq$ 2.5, 3.4 and 5.4   
    for $W_0/t\simeq$ 0.67, 1.0 and 2.0 
 while the present results show 
    $U_{c}/t \simeq 2.5$, 3.0 and 4.2 for $W_0/t=$ 0.67, 1.0 and 2.0 
      respectively. 
 A quantitative coincidence 
   between their  numerical calculations and ours are found 
    for $W_0/t \lsim 1$. 
 Further we note a fact deduced from the present result that  
   $\lan \cos \theta_{\rho +} \ran = 0$ 
  at $U = U_c$.
   Close to $U = U_c$,  the free energy 
  as a function of $\lan \cos \theta_{\rho +}\ran$ 
 takes a double minimum 
  for  the band  insulator 
 due to the degeneracy with respect to $\theta_{\rho +}=0$ and $\pi$ 
 while it takes a single minimum for the Mott insulator.    
 Thus the charge fluctuation 
  is strongly enhanced around  $U = U_c$ 
   as seen in the second order phase transition. 
  Such a  fact  could be related to 
 the enhancement of the dynamical charge\cite{Resta1} and  
  the increase of localization length \cite{Resta2} 
 at the critical point.

Finally we discuss about the effect of dimerization 
 where the corresponding Hamiltonian is  given by 
\begin{eqnarray}
 -  \td \sum_{j,\sigma} \,
      \left[  
         (-1)^j \, c_{j,\sigma}^{\dagger} c_{j+1,\sigma} 
                           \mbox{\rm h.c.} 
      \right]          
                 \point
         \label{eqn:4.1}
\end{eqnarray}
In terms of phase variable, eqs.~(\ref{eqn:field}) and (\ref{eq:thta}), 
 one can replace   eq.~(\ref{eqn:4.1})  as 
\begin{eqnarray}
\frac{4\td}{\pi \alpha}
\int \hspace{-1mm}\dx
    \sin \theta_{\rho +} 
    \cos \theta_{\sigma +} 
      \point 
         \label{eqn:td_phase}
\end{eqnarray}
 Equation (\ref{eqn:td_phase}) is compatible 
 with the $g_3$-term 
 in eq.~(\ref{eq:Hphase}) since the energy gain of the former 
   is obtained for 
     $\theta_{\rho +} =\pi/2$ and $\theta_{\sigma +} = \pi$
 (or $\theta_{\rho +} = -\pi/2$ and $\theta_{\sigma +} = 0$)
 while that of the latter is obtained for 
  $\theta_{\rho +} = \pm \pi/2$.
By noting that $\td$-term and the  $W_0$-term  
   have the same   periodicity with  $2 \pi$  as a function of  
      $\theta_{\rho + }$,   
  these two terms coexist for the relevant $G_{W}(l)$. 
In this case, the phase is determined so as to minimize 
 two energies of  $\td$-term and $W_0$-term. 
   We can incorporate eq.~(\ref{eqn:td_phase})
  into renormalization equations  
   and find 
 that $\td$-term is always relevant 
 and that there exists 
    a boundary between  the relevant $G_{W}(l)$ and 
     the irrelevant $G_{W}(l)$. 
 Based on these consideration, we can expect following. 
 Although  the qualitative feature is the same as that with $\td=0$,  
 there are noticeable facts that 
 the spin gap is strongly enhanced due to the relevant  $\td$
  and that  the critical value $U_c$ decreases 
   due to the enhancement of umklapp scattering by $\td$.

\section*{ Acknowledgements }
 This work was partially supported by  a Grant-in-Aid 
 for Scientific  Research  from the Ministry of Education, 
Science, Sports and Culture (Grant No.09640429).

\newpage

\appendix
\section{Derivation of the renormalization group equations}

\subsection{Coupling Constants}

The renormalization group equations are derived in a way similar to 
  Giamarchi and Schulz.\cite{Giamarchi_JPF}
By treating the nonlinear terms in eq.~(\ref{eq:Hphase})
  as the perturbation, 
  the response function, 
  $R_\rho (x,\tau) 
  \equiv \lan T_\tau \exp[\i\theta_{\rho +}(x,\tau)] 
                \exp[-\i\theta_{\rho +}(0,0)] \ran$,
  is calculated up to the third order as 
\begin{eqnarray}
R_\rho(\br_1-\br_2) &=& 
 \lan \e^{\i[\theta_{\rho +}(\br_1)-\theta_{\rho +}(\br_2)]} \ran_0
\nonumber \\ &+&
  \frac{1}{2}\frac{G_3^2}{(2\pi)^2}
  \sum_{\epsilon=\pm 1} 
  \int \frac{\d^2r_3}{\alpha^2}\frac{\d^2r_4}{\alpha^2}
    \left\{
          \lan \e^{\i[\theta_{\rho +}(\br_1)-\theta_{\rho +}(\br_2)
                      + \epsilon 2\theta_{\rho +}(\br_3)
                      - \epsilon 2\theta_{\rho +}(\br_4)
                     ]} 
          \ran_0
    \right.
\nonumber \\ && \hspace*{5cm}
    \left.
         -\lan \e^{\i[\theta_{\rho +}(\br_1)-\theta_{\rho +}(\br_2)]} 
          \ran_0
          \lan \e^{\i\epsilon[ 2\theta_{\rho +}(\br_3)
                              - 2\theta_{\rho +}(\br_4)
                     ]} 
          \ran_0
    \right\}
\nonumber \\ &+&
  \frac{1}{2}\frac{G_W^2}{(2\pi)^2}
  \sum_{\epsilon,\epsilon'=\pm 1} 
  \int \frac{\d^2r_3}{\alpha^2}\frac{\d^2r_4}{\alpha^2}
    \left\{
          \lan \e^{\i[\theta_{\rho +}(\br_1)-\theta_{\rho +}(\br_2)
                      + \epsilon \theta_{\rho +}(\br_3)
                      - \epsilon \theta_{\rho +}(\br_4)
                     ]} 
          \ran_0
    \right.
\nonumber \\ && \hspace*{1cm}
    \left.
         -\lan \e^{\i[\theta_{\rho +}(\br_1)-\theta_{\rho +}(\br_2)]} 
          \ran_0
          \lan \e^{\i\epsilon[ \theta_{\rho +}(\br_3)
                              - \theta_{\rho +}(\br_4)
                     ]} 
          \ran_0
    \right\}
    \lan \e^{\i\epsilon'[ \theta_{\sigma +}(\br_3)
                              - \theta_{\sigma +}(\br_4)]}\ran_0
\nonumber \\ &-&
\frac{1}{2}\frac{G_3 G_W^2}{(2\pi)^3}\sum_{\epsilon,\epsilon'} 
  \int \frac{\d^2r_3}{\alpha^2}\frac{\d^2r_4}{\alpha^2}
                              \frac{\d^2r_5}{\alpha^2}
    \left\{
          \lan \e^{\i[\theta_{\rho +}(\br_1)-\theta_{\rho +}(\br_2)
                      + \epsilon 2\theta_{\rho +}(\br_3)
                      - \epsilon  \theta_{\rho +}(\br_4)
                      - \epsilon  \theta_{\rho +}(\br_5)
                     ]} 
          \ran_0
    \right.
\nonumber \\ && \hspace*{0.5cm}
    \left.
         -\lan \e^{\i[\theta_{\rho +}(\br_1)-\theta_{\rho +}(\br_2)]} 
          \ran_0
          \lan \e^{\i\epsilon[ 2 \theta_{\rho +}(\br_3)
                              - \theta_{\rho +}(\br_4)
                              - \theta_{\rho +}(\br_5)
                     ]} 
          \ran_0
    \right\}
    \lan \e^{\i\epsilon'[ \theta_{\sigma +}(\br_4)
                              - \theta_{\sigma +}(\br_5)]}\ran_0
\nonumber \\ &-&
\frac{1}{2}\frac{G_1 G_W^2}{(2\pi)^3}\sum_{\epsilon,\epsilon'} 
  \int \frac{\d^2r_3}{\alpha^2}\frac{\d^2r_4}{\alpha^2}
                              \frac{\d^2r_5}{\alpha^2}
    \left\{
          \lan \e^{\i[\theta_{\rho +}(\br_1)-\theta_{\rho +}(\br_2)
                      + \epsilon  \theta_{\rho +}(\br_4)
                      - \epsilon  \theta_{\rho +}(\br_5)
                     ]} 
          \ran_0
    \right.
\nonumber \\ && \hspace*{0.5cm}
    \left.
         -\lan \e^{\i[\theta_{\rho +}(\br_1)-\theta_{\rho +}(\br_2)]} 
          \ran_0
          \lan \e^{\i\epsilon[   \theta_{\rho +}(\br_4)
                              - \theta_{\rho +}(\br_5)
                     ]} 
          \ran_0
    \right\}
    \lan \e^{\i\epsilon'[ 2\theta_{\sigma +}(\br_3)
                              - \theta_{\sigma +}(\br_4)
                              - \theta_{\sigma +}(\br_5)]}\ran_0
\nonumber \\ &+& 
    \cdots \virg
\label{eq:A1}
\end{eqnarray}
where 
 $\br=(x,\vf \tau)$ and 
$\langle \cdots \rangle_0$ denotes 
  an average  over the harmonic parts including $K_{\rho}$ or 
 $K_{\sigma}$ in eq.~(\ref{eq:Hphase}). 
 Scaling equations for  coupling constants
  up to the second-order can be calculated  from 
 response function, $R_\rho(r)$, which is expanded up to the 
  third-order for the nonlinear terms.
A straightforward calculation of eq.~(\ref{eq:A1}) yields,
\begin{eqnarray}
R_\rho(r_1-r_2) &=& 
 \e^{-K_\rho U(r_1-r_2)}
\nonumber \\ &+&
  \frac{1}{2}\frac{G_3^2}{(2\pi)^2}
  \sum_{\epsilon=\pm 1} 
  \int \frac{\d^2r_3}{\alpha^2}\frac{\d^2r_4}{\alpha^2}
    \e^{-K_\rho U(r_1-r_2)-4K_\rho U(r_3-r_4)}
\nonumber \\ && \times
    \left\{
     \e^{2\epsilon K_\rho [U(r_1-r_3)-U(r_1-r_4)-U(r_2-r_3)+U(r_2-r_4)]}
      -1
    \right\}
\nonumber \\ &+&
  \frac{G_W^2}{(2\pi)^2}
  \sum_{\epsilon=\pm 1} 
  \int \frac{\d^2r_3}{\alpha^2}\frac{\d^2r_4}{\alpha^2}
    \e^{-K_\rho U(r_1-r_2)-(K_\rho+K_\sigma) U(r_3-r_4)}
\nonumber \\ && \times
    \left\{
      \e^{\epsilon K_\rho [U(r_1-r_3)-U(r_1-r_4)-U(r_2-r_3)+U(r_2-r_4)]}
       -1
    \right\}
\nonumber \\ &-&
\frac{G_3 G_W^2}{(2\pi)^3}\sum_{\epsilon} 
  \int \frac{\d^2r_3}{\alpha^2}\frac{\d^2r_4}{\alpha^2}
                              \frac{\d^2r_5}{\alpha^2}
       \e^{-K_\rho U(r_1-r_2)+(K_\rho-K_\sigma) U(r_4-r_5)}
       \e^{-2K_\rho[U(r_3-r_4)+U(r_3-r_5)]}
\nonumber \\ && \times
   \left\{
      \e^{2\epsilon K_\rho [U(r_1-r_3)-U(r_2-r_3)]
         +\epsilon K_\rho [-U(r_1-r_4)-U(r_1-r_5)+U(r_2-r_4)+U(r_2-r_5)]
          } -1
   \right\}
\nonumber \\ &-&
\frac{G_1 G_W^2}{(2\pi)^3}\sum_{\epsilon} 
  \int \frac{\d^2r_3}{\alpha^2}\frac{\d^2r_4}{\alpha^2}
                              \frac{\d^2r_5}{\alpha^2}
       \e^{-K_\rho U(r_1-r_2)-(K_\rho-K_\sigma) U(r_4-r_5)}
       \e^{-2K_\sigma[U(r_3-r_4)+U(r_3-r_5)]}
\nonumber \\ && \times
   \left\{
      \e^{\epsilon K_\rho [U(r_1-r_4)-U(r_1-r_5)-U(r_2-r_4)+U(r_2-r_5)]
          } -1
   \right\} + \cdots \virg
\label{eq:A2}
\end{eqnarray}
  where $U(r)=\ln(r/\alpha)$ ($r=[x^2+(\vf \tau)^2]^{1/2}$). 
  We have discarded the difference of 
  velocities, i.e.,  $v_\rho=v_\sigma = \vf$.
  As for third-order with respect to coupling constant 
 in  eq.~(\ref{eq:A2}), 
 three cases  of
  $r_5=r_4+r$, $r_5=r_3+r$ and  $r_4=r_3+r$  with small $r$ 
 are chosen where 
   the expansion of these terms around $r=0$  is rewritten as
\begin{eqnarray}
&-&
\frac{G_3 G_W^2}{(2\pi)^3}\sum_{\epsilon} 
  \int \frac{\d^2r_3}{\alpha^2}\frac{\d^2r_4}{\alpha^2} \,
       \e^{-K_\rho U(r_1-r_2)-4K_\rho U(r_3-r_4)}
 \nonumber \\ && \times
   \left\{
      \e^{2\epsilon K_\rho [U(r_1-r_3)-U(r_1-r_4)
         -U(r_2-r_3)+U(r_2-r_4)]
          } -1
   \right\}
      \int \frac{\d^2r}{\alpha^2} \, \e^{(K_\rho-K_\sigma)U(r)}
\nonumber \\ &-&2
\frac{G_3 G_W^2}{(2\pi)^3}\sum_{\epsilon} 
  \int \frac{\d^2r_3}{\alpha^2}\frac{\d^2r_4}{\alpha^2}
       \e^{-K_\rho U(r_1-r_2)-(K_\rho+K_\sigma) U(r_3-r_4)}
\nonumber \\ && \times
   \left\{
      \e^{\epsilon K_\rho [U(r_1-r_3)-U(r_1-r_4)-U(r_2-r_3)+U(r_2-r_4)]
          } -1
   \right\}
      \int \frac{\d^2r}{\alpha^2} \, \e^{-2K_\rho U(r)}
\nonumber \\ &-&2
\frac{G_1 G_W^2}{(2\pi)^3}\sum_{\epsilon} 
  \int \frac{\d^2r_3}{\alpha^2}\frac{\d^2r_4}{\alpha^2}
       \e^{-K_\rho U(r_1-r_2)-(K_\rho+K_\sigma) U(r_3-r_4)}
\nonumber \\ && \times
   \left\{
      \e^{\epsilon K_\rho [U(r_1-r_3)-U(r_1-r_4)-U(r_2-r_3)+U(r_2-r_4)]
          } -1
   \right\}
      \int \frac{\d^2r}{\alpha^2} \, \e^{-2K_\sigma U(r)} 
+\cdots
\point
\label{eq:A3}
\end{eqnarray}
By comparing  eqs.~(\ref{eq:A2}) with  (\ref{eq:A3}), 
 it is found that 
  eq.~(\ref{eq:A3}) has an effect of the renormalization for 
 $G_3^2$-term and $G_W^2$-term   in eq.~(\ref{eq:A2}). 
Actually eq.~(\ref{eq:A2}) can be rewritten by the first three terms 
 with  effective coupling constants, which are given by 
\begin{eqnarray}
G_3^{{\rm eff} } &=& G_3 - G_W^2 \, 
      \int \frac{\d^2r}{\alpha^2} 
           \left(\frac{r}{\alpha}\right)^{1+K_\rho-K_\sigma}
\virg
\label{eq:A-G3}
\\
G_W^{{\rm eff}} &=& G_W 
   -  G_3\, G_W \, 
      \int \frac{\d^2r}{\alpha^2} 
           \left(\frac{r}{\alpha}\right)^{1-2K_\rho}
   - G_1\, G_W \, 
      \int \frac{\d^2r}{\alpha^2} 
           \left(\frac{r}{\alpha}\right)^{1-2K_\sigma}
\point
\label{eq:A-GW}
\end{eqnarray}
On the other hand, the effective quantity of $K_\rho$ can be 
 calculated from the response function up to the second order 
  for the nonlinear terms.  
 By expanding the exponential in  the second and the third terms 
  of eq.~(\ref{eq:A2}) 
 and using new variables  $\br=\br_3-\br_4$ and $\bR=(\br_3+\br_4)/2$, 
 eq.~(\ref{eq:A2})  is rewritten as,
\begin{eqnarray}
R_\rho(r_1-r_2) &=& \e^{-K_\rho U(r_1-r_2)}
\Bigl[
  1+
  2 \frac{G_3^2}{(2\pi)^2} K_\rho^2
     \, J_+(2K_\rho) \,I_+(r_1-r_2)
\nonumber \\ && \hspace{3cm}
+
  \frac{G_W^2}{(2\pi)^2} K_\rho^2
     \, J_+(K_\rho/2+K_\sigma/2) \,I_+(r_1-r_2)
 \Bigr] \virg
\label{eq:A6}
\end{eqnarray}
where
\begin{eqnarray}
I_+(r_1-r_2) &=& \int \d^2R \,\, U(r_1-R) \, \nabla_R^2 \, U(r_2-R)
= 2\pi \,U(r_1-r_2)
\virg
\\
J_+(K)&=& \int \frac{\d^2r}{\alpha^2} \, \frac{r^2}{\alpha^2} \,
    \e^{-2K U(r)}
 = 2\pi \int_\alpha^\infty \frac{\d r}{\alpha}
    \left(\frac{r}{\alpha}\right) ^{3-2K}
\point
\end{eqnarray}
 The scale invariance of  eq.~(\ref{eq:A6}) 
 is shown  
 by reexponentiating eq.~(\ref{eq:A6}) 
 where  the effective quantity of 
  $K_\rho$ is obtained as
\begin{eqnarray}
K_\rho^{\rm eff} &=& K_\rho -2 \, G_3^2 K_\rho^2 
   \int \frac{\d r}{\alpha} 
           \left(\frac{r}{\alpha}\right)^{3-4K_\rho}
-G_W^2 K_\rho^2 
   \int \frac{\d r}{\alpha} 
           \left(\frac{r}{\alpha}\right)^{3-K_\rho-K_\sigma}
\virg
\label{eq:A-Krho}
\end{eqnarray}
For the transformation given by $\alpha \to \alpha '=\alpha \e^{\d l}$,
  \cite{Giamarchi_JPF}
  these quantities have to be scaled as 
\begin{eqnarray}
K_\rho^{\rm eff}(K_\nu ',G_\mu';\alpha') &=& 
K_\rho^{\rm eff}(K_\nu ,G_\mu;\alpha)
\virg
\\
G_3^{{\rm eff} }(K_\nu ',G_\mu';\alpha') &=& 
G_3^{{\rm eff} }(K_\nu ,G_\mu ;\alpha) \, (\alpha'/\alpha)^{2-2K_\rho}
\virg
\\
G_W^{{\rm eff} }(K_\nu ',G_\mu';\alpha') &=& 
G_W^{{\rm eff} }(K_\nu ,G_\mu;\alpha) \,
     (\alpha'/\alpha)^{2-K_\rho/2-K_\sigma/2}
\virg
\end{eqnarray}
where 
  $K_\nu'$($\nu = \rho, \sigma$) and $G_\mu'$($\mu =3,1,W$) denote
  renormalized quantities.
 The quantity 
 $2K_\rho$ ($K_\rho/2+K_\sigma/2$) is a scaling dimension 
  for  $G_3^{{\rm eff}}$ (for $G_W^{{\rm eff}}$), which is obtained 
  from the second (third) term of eq.~(\ref{eq:A-Krho}). 
 By applying the infinitesimal transform to 
  eqs.~(\ref{eq:A-Krho}), (\ref{eq:A-G3}) and (\ref{eq:A-GW}),
 renormalized quantities are expressed as  
\begin{eqnarray}
K_\rho' &=& 
K_\rho -2 \, G_3^2 K_\rho ^2 \, \d l -G_W^2 K_\rho^2 \,\d l
\virg
\label{eq:A-Krho'}
\\
G_3' &=& G_3 + (2-2K_\rho) \,  G_3 \,\d l -G_W^2 \,\d l
\virg
\label{eq:A-G3'}
\\
G_W' &=& G_W + (2-K_\rho/2-K_\sigma/2) \, G_W \, \d l 
         -G_3\, G_W \, \d l -G_1\, G_W \, \d l
\virg 
\label{eq:A-GW'}
\end{eqnarray}
which lead to 
   eqs.~(\ref{eq:Krho}), (\ref{eq:G3}) and (\ref{eq:GW}), respectively.
 The renormalization group equations for 
  $K_\sigma(l)$ and $G_1(l)$ can be obtained 
 in a similar way to eqs.~(\ref{eq:Krho}) and  (\ref{eq:G3})
 by calculating
   the response function given by 
  $\lan T_\tau \exp[\i\theta_{\sigma +}(x,\tau)] 
                \exp[-\i\theta_{\sigma +}(0,0)] \ran$.
We note that, in case of $W_0/t=0$, these equations reduces to 
  to well known  equations 
 for half-filled Hubbard model.\cite{Solyom}

\subsection{Response Function}

 Based on the solution of  of eqs.~(\ref{eq:Krho})-(\ref{eq:GW}),
 we examine the response function, 
 $R_{\cos \theta_{\rho +}}(\br_1-\br_2) =
  \langle T_\tau 
     \cos \theta_{\rho+}(\br_1)  
     \cos \theta_{\rho+}(\br_2) 
  \rangle $,
 which   is calculated by writing 
  $R_{\cos \theta_{\rho +}} (r)=\exp[-K_\rho U(r)] 
                \cdot F_{\cos \theta_{\rho +}}(r)$.
 The  perturbative calculation for  $R_{\cos \theta_{\rho +}}$ 
 leads to 
\begin{eqnarray}
F_{\cos \theta_{\rho +}}(r_1-r_2) 
  &=&
  1
- \frac{G_3}{2\pi}\int\frac{\d^2r_3}{\alpha^2} 
   \e^{2K_\rho [U(r_1-r_2)-U(r_1-r_3)-U(r_2-r_3)]}
\nonumber \\ && \hspace{0cm}
-
   \left[ 
2 G_3^2
\int \frac{\d r}{\alpha}
    \left(\frac{r}{\alpha}\right) ^{3-4K_\rho}
+
G_W^2
\int \frac{\d r}{\alpha}
    \left(\frac{r}{\alpha}\right) ^{3-K_\rho-K_\sigma}
   \right] K_\rho^2 U(r_1-r_2) 
\point
\label{eq:A-pertub}
\end{eqnarray}
 The scaling invariance under the  transformation of 
    $\alpha \to \alpha '=\alpha \e^{\d l}$ is assumed for the 
 quantity $F_{\cos \theta_{\rho +}}(r)$, which  
 is expressed as 
\begin{eqnarray}
  F_{\cos \theta_{\rho +}}(r,K_\nu,G_\mu; \alpha) 
  = I_{\cos \theta_{\rho +}}(\d l,K_\nu,G_\mu) 
       \cdot F_{\cos \theta_{\rho +}}(r,K_\nu',G_\mu';\alpha')
\point
\label{eq:Ascale}
\end{eqnarray}
 The quantity  $I_{\cos \theta_{\rho +}}$ is 
    determined perturbatively.
By applying the infinitesimal transform to eq.~(\ref{eq:A-pertub}),
  we obtain
\begin{eqnarray}
                    \label{eq:aa}
F_{\cos \theta_{\rho +}}
  &=&
  1 -2 \, G_3 \, \d l 
   + ( 2 \, G_3^2 + G_W^2) \, K_\rho^2 \, U(r_1-r_2) \, \d l
\nonumber \\ && \hspace{0cm}
- \frac{G_3'}{2\pi}\int_{\alpha'} \frac{\d^2r_3}{\alpha'^2} 
   \e^{2K_\rho [U'(r_1-r_2)-U'(r_1-r_3)-U'(r_2-r_3)]}
\nonumber \\ && \hspace{0cm}
-
   \left[ 
2 G'^2_3
\int_{\alpha'}^{\infty} \frac{\d r}{\alpha'}
    \left(\frac{r}{\alpha'}\right) ^{3-4K_\rho}
+
G'^2_W
\int_{\alpha'}^{\infty} \frac{\d r}{\alpha'}
    \left(\frac{r}{\alpha'}\right) ^{3-K_\rho-K_\sigma}
   \right] K_\rho^2 \, U'(r_1-r_2) 
\virg
\end{eqnarray}
where $U'(r) = \ln (r/\alpha')$. 
 From eq.~(\ref{eq:aa}), 
  the multiplicative factor $I_{\cos \theta_{\rho +}}$ is obtained as
\begin{eqnarray}
I_{\cos \theta_{\rho +}}(\d l,K_\nu,G_\mu)  &=& 
\exp \left[
        -2 \, G_3 \, \d l + 2 \, G_3^2 \, K_\rho^2 \, U(r) \, \d l
	             +  G_W^2 \, K_\rho^2 \, U(r) \, \d l
     \right]
\point
\end{eqnarray}
 The transformation of eq.~(\ref{eq:Ascale}) is performed recursively
  until the new cutoff $\alpha'$ reaches $r$. 
 Thus  the  reconstructed  $F_{\cos \theta_{\rho +}}$ is expressed as
\begin{eqnarray}
F_{\cos \theta_{\rho +}}(r,K_\nu,G_\mu;\alpha)  &=& 
\exp \left[ \int_0^{\ln (r/\alpha)} \d l \, 
         \ln I_{\cos \theta_{\rho +}}(\d l,K_\nu(l),G_\mu(l))
     \right]
\point
\end{eqnarray}
 The terms including the second order of the 
  coupling constants are rewritten as follows. 
From eq.~(\ref{eq:Krho}), one obtains,
\begin{eqnarray}
&&
\int_0^{\ln(r/\alpha)} \d l
       \left[  
                 2  \, G_3^2(l) \, K_\rho^2(l)
	             +  G_W^2(l) \, K_\rho^2 (l)
       \right] \ln\left[\frac{r}{\alpha(l)}\right]
\nonumber \\
&&= K_\rho (0) \ln(r/\alpha) - \int_0^{\ln(r/\alpha)} \d l \,K_\rho (l),
\end{eqnarray}
  where $\alpha(l)=\alpha \, \e^{l}$. 
Thus the response function $R_{\cos \theta_{\rho +}}$ is expressed as
\begin{eqnarray}
R_{\cos \theta_{\rho +}}(r) 
&=&
\exp \left[- \int_0^{\ln(r/\alpha)} \d l \,\,
  \bigl(K_\rho (l) +2G_3(l) \bigr)\right]
\end{eqnarray}
which leads to eq.~(\ref{eq:response_r}).
In a similar way, 
 another  response function 
leads to eqs.~(\ref{eq:response_s}).

\section{Evaluation of  cutoff parameter}

We calculate response function of charge density wave 
 with  $Q=2\kf +q$ and  $\kf=\pi/(2a)$, which is expressed as  
\begin{eqnarray}
              \label{eq:PAI}
\Pi(Q)= 
\frac{1}{2} \sum_{j,\sigma} \int \d\tau \, 
  \left.
  \lan T_\tau c_{j,\sigma}^\dagger (\tau) \, 
              c_{j,\sigma} (\tau) \,
              c_{0,\sigma}^\dagger (0) \, 
              c_{0,\sigma} (0) \ran
       \e^{-\i QR_j+\i\omega_m \tau} \right|_{\i \omega_m \to 0}
\point
\end{eqnarray}
  For the case with only the first term of eq.~(\ref{eq:H}),
 eq.~(\ref{eq:PAI}) is calculated as 
\begin{eqnarray}
\Pi(Q)
&=&
\frac{1}{2\pi} \int_{-\pi/a}^0 \d k \, 
  \frac{f(\varepsilon_{k+Q})-f(\ek)}{\ek - \varepsilon_{k+Q}}
\,\,\,  \too_{qa \ll 1} \,\,\,
 = \frac{1}{2\pi\vf} \ln \frac{C}{qa}
\virg
\label{eq:B-fermion}
\end{eqnarray}
  where $C \simeq 1.273 \pi$.
 The evaluation by the bosonization method leads to \cite{Solyom}
\begin{eqnarray}
\Pi_{\rm B}(Q)
&=&
\frac{1}{(2\pi\alpha)^2} \int \dx  \d\tau \left. 
             \frac{1}{1+\left(\vf \tau/\alpha\right)^2 
                       +\left(x/\alpha\right)^2} 
      \e^{\i qx-\i\omega_m \tau} \right|_{\i\omega_m \to 0}
\nonumber \\
&=& \frac{K_0(q\alpha)}{2\pi\vf} 
  \,\,\,  \too_{q\to 0} \,\,\,
 \frac{1}{2\pi\vf} \, \ln\frac{1}{q\alpha}
\point
\label{eq:B-boson}
\end{eqnarray}
From eqs.~(\ref{eq:B-fermion}) and (\ref{eq:B-boson}), 
  it is found that
$ \alpha \simeq  a/(1.273 \pi)$.

\end{document}